\documentclass[fleqn,usenatbib]{mnras}
\usepackage{newtxtext,newtxmath}
\usepackage{bm} 
\usepackage{dcolumn}
\usepackage{dsfont}
\usepackage[T1]{fontenc}
\usepackage{ae,aecompl}
\usepackage{graphicx}	
\usepackage{amsmath}	
\usepackage{amssymb}	
\usepackage{multirow}
\usepackage{hyperref}

\title[Localised acceleration]
  {A method to calculate gravitational accelerations within discrete localised regions in the Milky Way}
\author[Rain Kipper, Elmo Tempel and Peeter Tenjes]{Rain Kipper$^{1}$\thanks{E-mail:
rain.kipper@ut.ee}, Elmo Tempel$^{1,2}$ and Peeter Tenjes$^{1}$\\
$^1$Tartu Observatory, University of Tartu, 61602 T\~oravere, Tartumaa, Estonia\\
$^{2}$Leibniz-Institut f\"ur Astrophysik Potsdam (AIP), An der Sternwarte 16, 14482 Potsdam, Germany
}
\date{Accepted X. Received X; in original form X}

\pagerange{\pageref{firstpage}--\pageref{lastpage}} \pubyear{2018}

\def\LaTeX{L\kern-.36em\raise.3ex\hbox{a}\kern-.15em
    T\kern-.1667em\lower.7ex\hbox{E}\kern-.125emX}

\begin{document}

\label{firstpage}

\maketitle

\begin{abstract}

We present a method to calculate gravitational potential gradients within regions containing few tens of thousands stars with known phase space coordinates. 
The central idea of the method is to calculate orbital arcs for each star within a given region for a certain parametrised  potential (gravitational acceleration) and to assume that position of each star on its orbital arc is a random variable with a uniform probability density in time. Thereafter, by combining individual probability densities of stars it is possible to calculate the overall probability density distribution and likelihood for a given region as a function of gravitational acceleration parameters. The likelihood has a maximum if the calculated probability distribution and the observed distribution of stars in phase space are consistent. This allows us to constrain gravitational accelerations and potential gradient values. The method assumes that phases of stars are mixed within the regions where stellar orbits are calculated. We tested the method for 12 small rectangular regions within simulated disc galaxy from Gaia Wiki. Tests show that even with a rather simple acceleration form the calculated accelerations in galactic plane coincide with their true values from simulation about 5 per cent, misalignment between the calculated and true acceleration vector directions is less than 1 degree (median values). The model can be used with the Milky Way Gaia complete solution data. 
\end{abstract}
\section{Introduction} 
\label{sec:introduction}

One of the most important functions determining kinematics and dynamics of galaxies is the gradient of the gravitational potential. When calculating gravitational potential and/or their derivatives, it is usually assumed that a galaxy is in a stationary state and its gravitational potential has some symmetry. Starting from the surface brightness distribution a galaxy is assumed to consist of one or several axisymmetric or spherical components, the luminosity density distribution of which being described with a sufficiently simple analytical formula. Then, assuming that a similar formula is valid also for the mass density distribution it is possible to calculate gravitational potential derivatives by solving, for example, Jeans equations \citep[e.g.][]{Binney:1990, vdMarel:1990, Emsellem:1994, Cappellari:2008}. Free parameters in the density distribution formulae can be derived by comparing calculated model predictions with the spectral and kinematic data of the galaxy. These kind of models can be rather simple (e.g. two-component models) and are thus suitable to model a large number of galaxies \citep[see e.g.][]{Mcgaugh:2016, Kalinova:2017, Barone:2018, Li:2018}. 

More detailed but also more time consuming methods are for example models using Schwarzschild orbit-based method \citep{Schwarzschild:1979,  Thomas:2004, Valluri:2004, Cappellari:2006, Cappellari:2007, vdven:2008, Kowalczyk:2017} or action-angle torus-mapping technique \citep{McGill:1990, Copin:2000, Bovy:2014, Binney:2015, Binney:2016}. These models can be rather flexible, that is, with a sufficient number of components it is possible to describe a galaxy with nearly arbitrary density distributions. These models take into account three integrals of motion and allow to describe galactic components with three-axial ellipsoidal density distributions and even bars and spiral arms \citep{Binney:2018}. Both methods assume global stationarity of a galaxy as complete orbits are calculated there. Unfortunately, corresponding calculations are more time consuming and the number of galaxies modelled with these methods is limited\footnote{However, surprisingly large number of galaxies have been modelled within Califa project \citep{Zhu:2018}.}.

Modelling of individual galaxies via N-body simulations (e.g. made to measure) is usually free of symmetry and stationarity assumptions but these models are even more time consuming and their spatial and mass resolution is yet moderate \citep{Syer:1996, deLorenzi:2007, Long:2010, Zhu:2014}.

Several galactic main components (stellar disc, bulge, stellar halo) often seem to be symmetric. But knowing the importance of interactions in galaxy evolution it is not always clear how good symmetry assumptions really are. As an example, for our Milky Way (MW) galaxy, it is not clear how important is the central bar while modelling the overall gravitational potential \citep[e.g.][]{Binney:2018}. 

An excellent opportunity to study gravitational potential of the MW is given by the Gaia data \citep{Gaia_DR2}. To calculate accelerations directly for individual stars from Gaia data requires very precise full six-dimensional phase space information, which is not available in Gaia. However, there is a possibility to calculate acceleration components from Gaia data within a sufficiently small regions containing many (several thousands) stars. In present paper we propose a method to do this and we test the method by using simulated data.  

Application of the proposed method assumes that Gaia complete solution data is available, i.e one has  sufficiently precise 3D coordinates and velocities for a large number of stars. By selecting a region around a point in a galaxy containing several tens of thousands stars and calculating orbits of these stars it is possible to calculate gravitational potential derivatives at a given point rather precisely. This paper continues our previous work \citep{Kipper:2018} where only Solar neighbourhood in 1D was studied. With Gaia complete solution data the entire MW covered by sufficiently precise Gaia data can be analysed.

The structure of the paper is following. In Section~\ref{sec:method} we describe our proposed method to calculate the gravitational acceleration within discrete localised regions. In Section~\ref{sec:testing_the_method} we apply the method to simulated data taken from Gaia Challenge project. And finally, in Section~\ref{sec:results} and Section~\ref{sec:summary} we present our results and conclude the paper, accordingly.

\section{Method} 
\label{sec:method}

In this Section we present a method allowing to calculate gravitational acceleration  vector components in a small bounded region $D$ that contains sufficient number of stars with known phase space coordinates. These individual regions should be taken sufficiently small in order to make an assumption that gravitational potential derivatives within these region can be approximated with a simple analytical form. No assumptions about symmetries of the overall mass distribution are necessary. 

The whole gravitational acceleration field of a galaxy can be calculated by selecting many regions of this kind all over the galaxy. 

We assume that phases of stars are mixed only within these individual regions $D$, that is, they are locally mixed. This is different from the assumption of completely mixed phases.

Let us consider first an orbit (a set of similar valued integrals of motion) extending all over the galaxy and containing sufficiently large number of stars. Let us describe positions of a star within this orbit as phases $\tau_i$ on the orbit. If the phases are completely mixed the distribution of $\tau_i$ is uniform, or equivalently, the phase distances between stars on the orbit are taken from exponential distribution with the mean value of $\Delta\tau$.

If the phases are not completely mixed but only moderately mixed (corresponding to a somewhat non-stationary galaxy) then $\Delta\tau$ slightly changes along the orbit. In this case we may introduce a concept of locally mixed phases, i.e. globally $\Delta\tau$ values can be different, but within a sufficiently small arc of the orbit $\Delta\tau$ can be approximated as $\Delta\tau \simeq \mathrm{const}$. For different arcs (or different regions $D$) this constant is different. Thus, the method allows that for a whole galaxy the phases of stars in a given orbit are not completely mixed (e.g. near to a bar). However, by selecting a sufficiently small region $D$, one can approximate the phases to be mixed locally.

Knowing the phase space coordinates of a star $i$ in a particular moment $t_0$ $(\textbf{\textit{x}}_i (t_0)$, $\textbf{\textit{v}}_i (t_0))$ and accelerations $\textbf{\textit{a}}(\textbf{\textit{x}}_i)$ caused by an external potential it is possible to integrate the equations of motion for the star $i$ within a region $D$  
\begin{eqnarray}
	\frac{\mathrm{d}\textbf{\textit{x}}_i}{\mathrm{d}t} & = & \textbf{\textit{v}}_i(t)\label{eq:liikumisvorrandid1} \\
	\frac{\text{d}\textbf{\textit{v}}_i}{\mathrm{d}t} & = &\textbf{\textit{a}} (\textbf{\textit{x}}_i).\label{eq:liikumisvorrandid2}
\end{eqnarray}
Here $t$ is time and also a parameter that determines the position of a star in an orbit. Initial conditions $\textbf{\textit{x}}_i (t_0)$ and $\textbf{\textit{v}}_i (t_0)$ can be taken e.g. from Gaia observations. 

For a given acceleration $\textbf{\textit{a}}(\textbf{\textit{x}}_i)$ a solution ($\textbf{\textit{x}}_i(t)$, $\textbf{\textit{v}}_i(t)$) can be derived. Next we use  $\textbf{\textit{q}}$ as a shorter designation of all six phase space coordinates. For all stars orbits are calculated within a region $D$ and we assume that within this region acceleration components in Eq.~(\ref{eq:liikumisvorrandid2}) are only functions of coordinates or can be approximated within $D$ as such.

Let $p_i (\textbf{\textit{q}}_i(t))\, \text{d}t$ be the probability to find a star $i$ at its orbital point $\textbf{\textit{q}}_i (t)$ within an interval determined by $t$ and $t + \text{d}t$. For all stars these probabilities can be calculated by solving equations of motion (\ref{eq:liikumisvorrandid1}) and (\ref{eq:liikumisvorrandid2}), and then calculating how much time a star spends at these phase space coordinate intervals. 

The probability to find a star near $\textbf{\textit{q}}$ is an integral of corresponding probability density over the time star spends at corresponding orbital arc $\int p_i (\textbf{\textit{q}}_i) \text{d}t.$ If we consider the whole region $D$ this probability should be normalised to one
\begin{equation}
\int\limits_{\textbf{\textit{q}}_i\in D} p_i(\textbf{\textit{q}}_i(t))\text{d}t = 1. \label{eq:int}
\end{equation}

In case of a stellar system, where the phases of stellar orbits are randomised, the probability to find a star at a specific phase on its orbit does not depend on time
\begin{equation}
\frac{\text{d}p_i( \textbf{\textit{q}}_i (t) )}{\text{d}t} = 0. \label{eq:tul}
\end{equation} 
Thus, taking into account Eqs.~(\ref{eq:int}) and (\ref{eq:tul}) we have that the probability density in (\ref{eq:int}) can be written in form
\begin{equation}
	p_i( \textbf{\textit{q}}_i (t) ) = \text{const} = \frac{1}{t_\text{max}-t_\text{min}}.  \label{eq:pi_vorrand}
\end{equation}
Here coordinates and velocities are taken along the orbit of a star $i$ and $t_\text{min}$ and $t_\text{max}$ indicate the times when this star enters and leaves the region $D$. As we assumed that phases of stars are mixed only within a region $D$, for different regions the constant in Eq.~(\ref{eq:pi_vorrand}) can be different.

Combining individual stellar probability distribution functions of stars, we can calculate the overall probability distribution function at all points $\textbf{\textit{q}}$ in $D$
\begin{equation}
p(\textbf{\textit{q}}) = \frac{1}{U} \sum\limits_{i}\int p_i(\textbf{\textit{q}}_i(t))\, K(\textbf{\textit{q}}_i(t), \textbf{\textit{q}})\, \text{d}t. \label{eq:kernel}
\end{equation}
Here $U$ is a normalising constant, summation is over the orbits of all stars and integral is taken over the time a star spends in $D$. The kernel function $K$ is needed to convert the orbital arcs into contiguous six-dimensional function by summing over all orbits. A simple example of the kernel function is a top-hat function describing whether an orbital arc $\textbf{\textit{q}}_i$ is near to a particular point $\textbf{\textit{q}}$.

For a given gravitational acceleration form, we maximise the likelihood function (or likelihood multiplied by a prior when using a Bayesian analysis) to find the best parameters for the acceleration. Likelihood is constructed using observations and probability densities from Eq.~(\ref{eq:kernel}) as follows
\begin{equation}
	\mathcal{L} = \prod\limits_k p (\textbf{\textit{q}}_k) \quad \mathrm{or} \quad \log \mathcal{L} = \sum\limits_k \log p (\textbf{\textit{q}}_k). \label{eq:likelihood}
\end{equation}
The maximum of this function gives the values to a priori best-fit acceleration function $\textbf{a}(\textbf{x})$
\begin{equation}
\textbf{a}(\textbf{x}) = \operatorname*{arg\,max}_{a} \mathcal{L},
\end{equation}
where the maximum is taken over all possible values (e.g. using a fixed parametrised form) of $\textbf{\textit{a}}$.

\section{Testing of the method on simulation} 
\label{sec:testing_the_method}

\begin{figure*}	\includegraphics{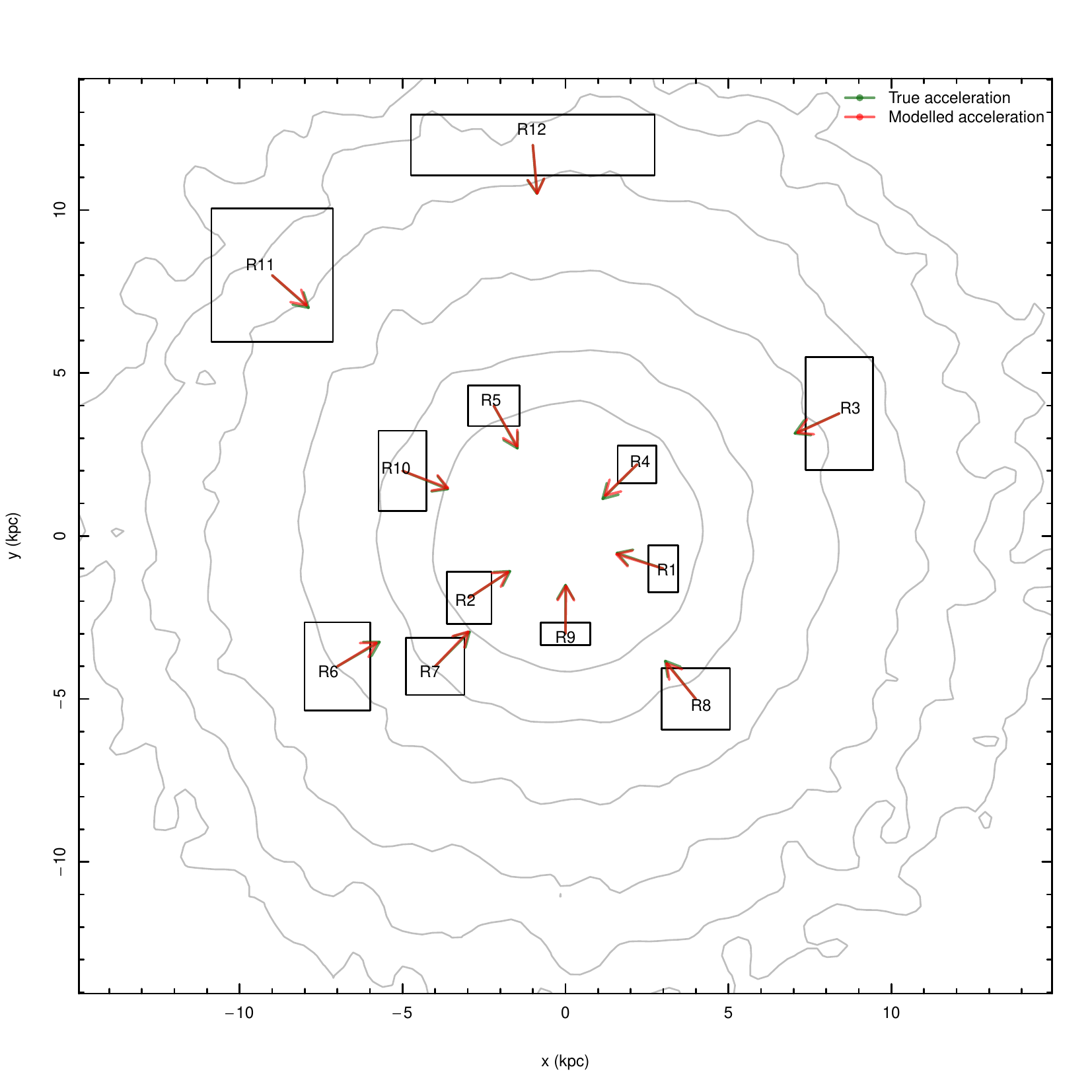}
	\caption{Calculated (red arrows) and true (green arrows) acceleration vectors in the plane of the simple snapshot of the simulated galaxy. True accelerations are normalised to constant length and simulated ones according to them. We note that some of the red and green arrows are exactly on top of each other. Simulated regions together with their names are presented as black rectangular boxes (see also Table~\ref{tab:regions}). 
}\label{fig:gal_plane_NE}
\end{figure*}
\begin{figure*}
	\includegraphics{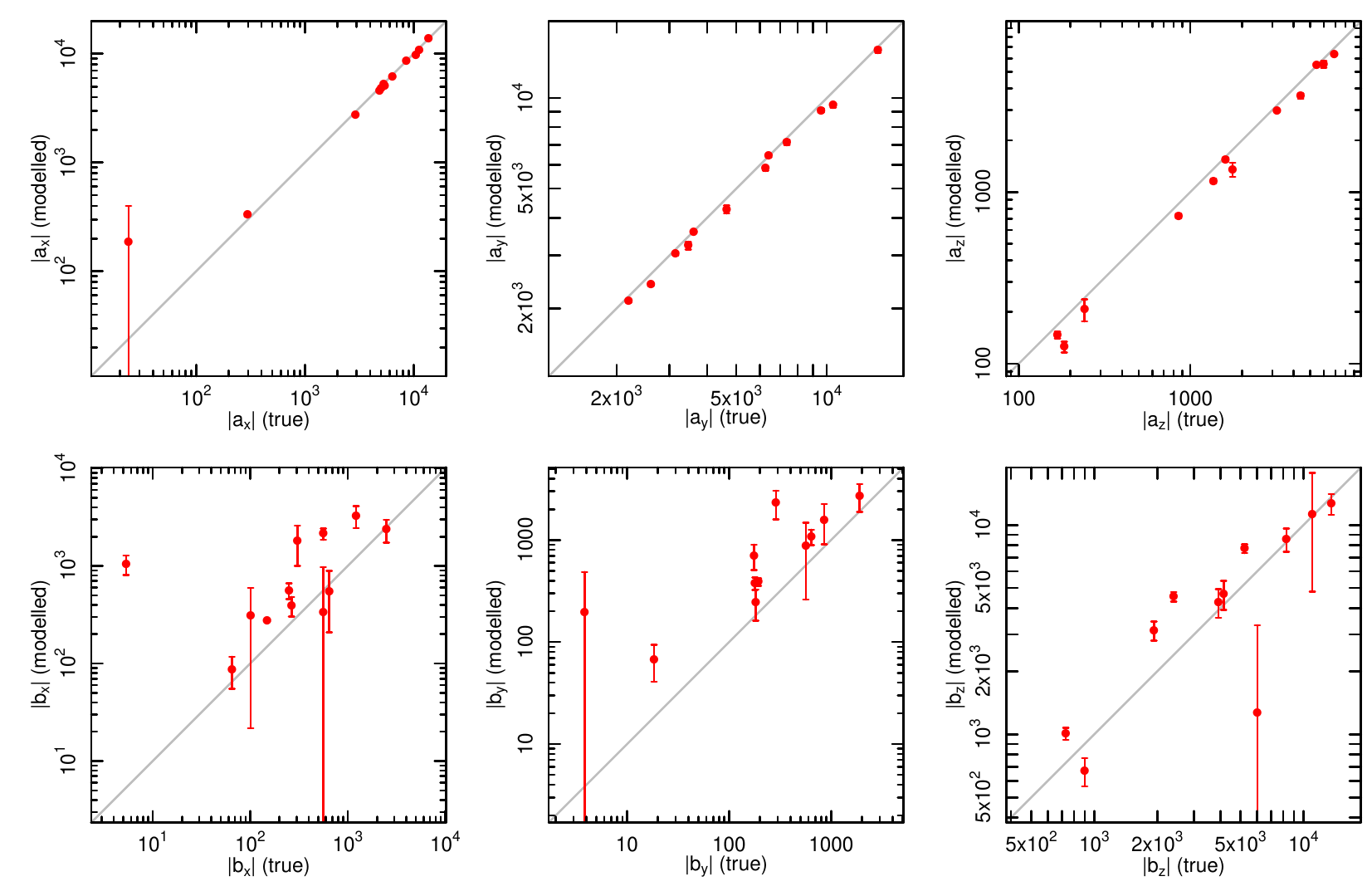}
	\caption{Modelled and true parameters together with error bars of accelerations (see Eq.~(\ref{eq:model_acc})) for twelve regions in the simple snapshot of the simulation. The errors are marginalised standard deviations of the posterior distribution samples. It is seen that acceleration components are rather well recovered (upper panels), but correction terms for accelerations for several regions have rather large uncertainties. } \label{fig:parval_NE}
\end{figure*}
\begin{figure*}
	\includegraphics{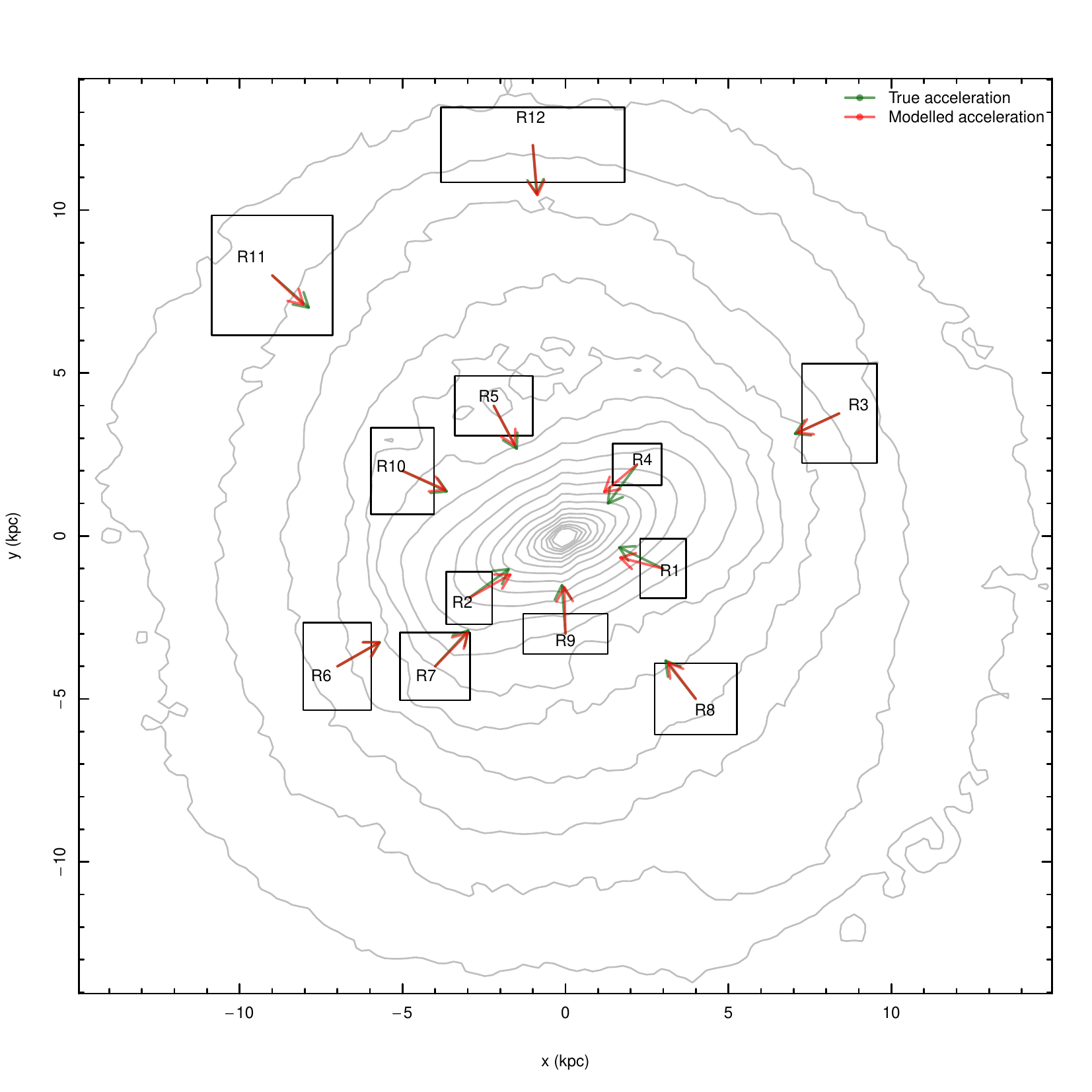}
	\caption{Calculated (red arrows) and true (green arrows) acceleration vectors in the plane of the barred snapshot of the simulated galaxy. True accelerations are normalised to constant length and simulated ones scaled according to them. Simulated regions together with their names are presented as black rectangular boxes (see Table~\ref{tab:regions} for numerical accuracies). 
}\label{fig:gal_plane}
\end{figure*}

To test the method, we chose two galaxy snapshots taken from the simulation \citep[see][]{Garbari:2011} published in the Gaia Challenge wiki\footnote{\url{http://astrowiki.ph.surrey.ac.uk/dokuwiki/doku.php?id=tests:discs:mockreadfullnbm}}. 
Initial conditions of the simulation correspond to conditions giving a system close to equilibrium. Simulated galaxy consists of three components -- the bulge, the disc and the dark matter halo. One snapshot was taken at $0.049$ Gyr from the start of the simulation and corresponds to a nearly axisymmetric galaxy. We denote this snapshot as a simple  galaxy\footnote{In the Gaia wiki, this snapshot is referred as unevolved snapshot.}. The second snapshot was taken from $4.018$ Gyr from the beginning of the simulation and is similar to a barred disc galaxy. We denote this snapshot as a barred galaxy. Surface density isophotes of these two snapshot mocks are seen in Figs.~\ref{fig:gal_plane_NE} (a simple galaxy) and \ref{fig:gal_plane} (a barred galaxy). 

In case of the second snapshot due to the presence of the bar,  accelerations in Eq.~(\ref{eq:liikumisvorrandid2}) are probably also functions of time. This snapshot is used to check sensitivity of the method to non-stationarity effects and to test the method in a more realistic situation.

Regions $D$ used for calculations may have rather different shapes (box, ellipsoid, cone etc) and sizes. To calculate $p$ with a sufficient accuracy, these regions should contain a sufficient number of stars. In addition, as our intention is to determine acceleration of stars but since acceleration acts slowly (it takes time or needs a sufficient distance) these regions need to be large enough. 

Simulation results used to test the method are given in cartesian coordinates. Due to this we selected regions $D$ as rectangular boxes with sides aligned with the coordinate axes. Relative proportions of  boxes were chosen from the criteria that in all directions for fast stars (mean velocity + two standard deviations) it would take about the same time to pass through the box. Overall dimensions of regions were chosen by demanding that regions should contain sufficient number of stars.

We selected twelve different regions. All regions contain $100\,000 \pm 20\,000$ stars. Since we tested our results also for shot noise with smaller samples and in order to keep the tests and results consistent, all these regions were modelled with $50\,000$ stars. Locations of regions were chosen to include most relevant regions in the test galaxies: intermediate "Solar neighbourhood" regions, outer disc regions and regions surrounding the central bulge or bar. In vertical $z$ direction centres of the regions were selected to lie outside of the galactic plane to  test the method in various conditions. For both snapshots regions positions were the same, although due to slightly different kinematical properties their sizes were slightly different. These twelve regions superposed to the surface density isophotes for two snapshot mocks are shown in Figs.~\ref{fig:gal_plane_NE} and \ref{fig:gal_plane} as rectangular boxes.

Analytical form for the acceleration $\textbf{\textit{a}}(\textbf{\textit{x}})$ (or gravitational potential) within $D$ is rather free, that means it may contain quite a lot of free parameters (e.g. Taylor series with many terms). In principle, this sets a limit to the maximum size of a region. However, these limiting conditions for gravitational potential are rather weak.

In present paper, acceleration components are taken to be linear functions of the coordinates
\begin{equation}
	a_j = a_{0,j} + b_{j}\cdot(x_j - x_{\mathrm{0},j}). \label{eq:model_acc}
\end{equation}
Here $j$ indices cartesian coordinates, $a_{j,0}$ is the acceleration at the centre of the region, $x_{\mathrm{0},j}$ and $b_j$ are linear term parameters. This form (compared to e.g. $a_j = a_{0,j}+b_jx_j$) helps to keep the model parameters $a_j$ and $b_j$ a priori maximally independent, which is advantageous during modelling, and keeps the interpretation of $a_j$ as the acceleration in the centre of the box. 
In case of $z$ component it is possible to use an assumption of intrinsic symmetry of a galaxy with respect to the galactic plane and to write $z$ component of Eq.~(\ref{eq:model_acc}) in a different form: $a_z = b_z\cdot |z - z_0|$. In the present paper for all components we used a general form (\ref{eq:model_acc}). In this case for every region we have six free parameters to fit. 

To test the accuracy of the model, true accelerations were calculated directly by using least square regression to the gravitational potential values $a_j = -\partial\Phi/\partial x_j$ at the locations of simulation particles. Regression details (e.g. whether we use all the stars from the region to calculate potential gradient or only a subsample of them, whether to add mixed terms to the $a_j$ form etc) introduce uncertainties to the true acceleration values up to few hundred km$^2$s$^{-2}$kpc$^{-1}$. This is the maximum accuracy we may expect from our tests. 

For the kernel function $K$ in Eq.~(\ref{eq:kernel}) we adopted a gridding approach: in a particular region $D$ a six-dimensional grid was constructed and for each orbit the stellar orbital probabilities were summed within a grid cell 
\begin{equation}
	K ( \textbf{\textit{q}}_i, \textbf{\textit{q}} ) = \mathds{1}(\{\textbf{\textit{q}}_i, \textbf{\textit{q}} \} \in g).
\end{equation}
Here $\mathds{1}$ is an indicator function describing whether $\textbf{\textit{q}}_i$ and $\textbf{\textit{q}}$ are in the same grid-cell $g$. In this case $p( \textbf{\textit{q}} )$ will be a sum of top-hat functions. This approach does not widen probability density profile perpendicular to the orbit but bins, hence there will be no artificial distortion to the $p$. Number of grid cells for all dimensions were taken $6$ thus the total number of cells in a particular region is rather large (about 47000) and in average there was one star per cell. 

For the fitting process, we used Multinest code for the Bayesian analysis \citep{MN1, MN2,MN3}. It is a well tested code that is aimed to converge to the solution in cases of multimodal posterior distributions. The selected prior distributions were uniform with rather wide limits $-28\,000\dots28\,000$ km$^2$s$^{-2}$kpc$^{-1}$ for $a_{j,0}$ and $-28\,000\dots28\,000$ km$^2$s$^{-2}$kpc$^{-2}$ for $b_{j}$. We used 5000 live points during the fitting, and adopted the mean of the equal weight output posterior samples as our modelled values.

\section{Results} 
\label{sec:results}

    	\begin{table*}
		\caption{General parameters of selected regions and accuracies of the calculated accelerations. The first columns give the name of the region and galactocentric cylindrical coordinates ($R$ and $z$) of the regions centres. The rest of the columns are divided according to the test snapshots, both having four parameters: $\Delta z/2$ as semi-height of the modelled region,  $a_\text{R}$ as planar acceleration relative error ((true value - modelled value)/true value), $a_z$ as vertical acceleration relative error, and directional difference between modelled and true acceleration vectors. }\label{tab:regions}
		\begin{tabular}{lcc|cccc|cccc}
\hline\hline
			& & & \multicolumn{4}{c|}{Simple galaxy} & \multicolumn{4}{c}{Barred galaxy} \\
			\!\!Name\!\!  &  R  & z &  \multicolumn{1}{c}{$\Delta z/2$} & \multicolumn{1}{c}{$a_\text{R}$}  &  \multicolumn{1}{c}{$a_z$} &   \multicolumn{1}{c|}{\!\!Direction\!\!} & \multicolumn{1}{c}{$\Delta z/2$} & \multicolumn{1}{c}{$a_\text{R}$}  &  \multicolumn{1}{c}{$a_z$} &   \multicolumn{1}{c}{\!\!Direction\!\!} \\
			\hline
			 & kpc &  kpc  & kpc &  \% & \% & deg & kpc &  \% & \% & deg \\
			\hline
R1 &  3.13  &  0.2  &  0.2  &  -0.13  &  -1.1  &  1.6  &  0.28  &  9.04  &  23.87  &  11.4 \\ 
R2 &  3.47  &  0.34  &  0.23  &  4.3  &  7.84  &  0.4  &  0.25  &  3.07  &  -1.93  &  6.5 \\ 
R3 &  9.07  &  -0.06  &  0.2  &  5.72  &  14.5  &  0.6  &  0.21  &  5.89  &  63.6  &  0.5 \\ 
R4 &  3.1  &  -0.22  &  0.18  &  8.57  &  7.45  &  0.9  &  0.23  &  11.92  &  16.06  &  13.1 \\ 
R5 &  4.62  &  0.1  &  0.18  &  3.98  &  23.24  &  1.1  &  0.25  &  8.64  &  10.68  &  0.5 \\ 
R6 &  7.99  &  0.15  &  0.21  &  6.31  &  15.22  &  1.2  &  0.21  &  1.63  &  0.48  &  0.5 \\ 
R7 &  5.58  &  -0.15  &  0.19  &  1.66  &  3.82  &  1.3  &  0.2  &  4.36  &  22.83  &  1.5 \\ 
R8 &  6.38  &  -0.15  &  0.21  &  5.88  &  15.12  &  0.7  &  0.22  &  6.36  &  6.36  &  1.1 \\ 
R9 &  3.08  &  -0.15  &  0.21  &  2.24  &  17.64  &  0.6  &  0.31  &  4.66  &  -0.16  &  2.7 \\ 
R10 &  5.39  &  -0.3  &  0.23  &  0.78  &  7.08  &  1.3  &  0.29  &  4.88  &  7.06  &  1.2 \\ 
R11 &  11.76  &  0.1  &  0.22  &  6.69  &  32.01  &  0.7  &  0.22  &  12.27  &  4.81  &  1.7 \\ 
R12 &  11.92  &  0.11  &  0.3  &  0.41  &  12.51  &  0.5  &  0.23  &  -3.18  &  18.05  &  0.8 \\ 
\hline
		\end{tabular}
	\end{table*}

\begin{figure*}
	\includegraphics{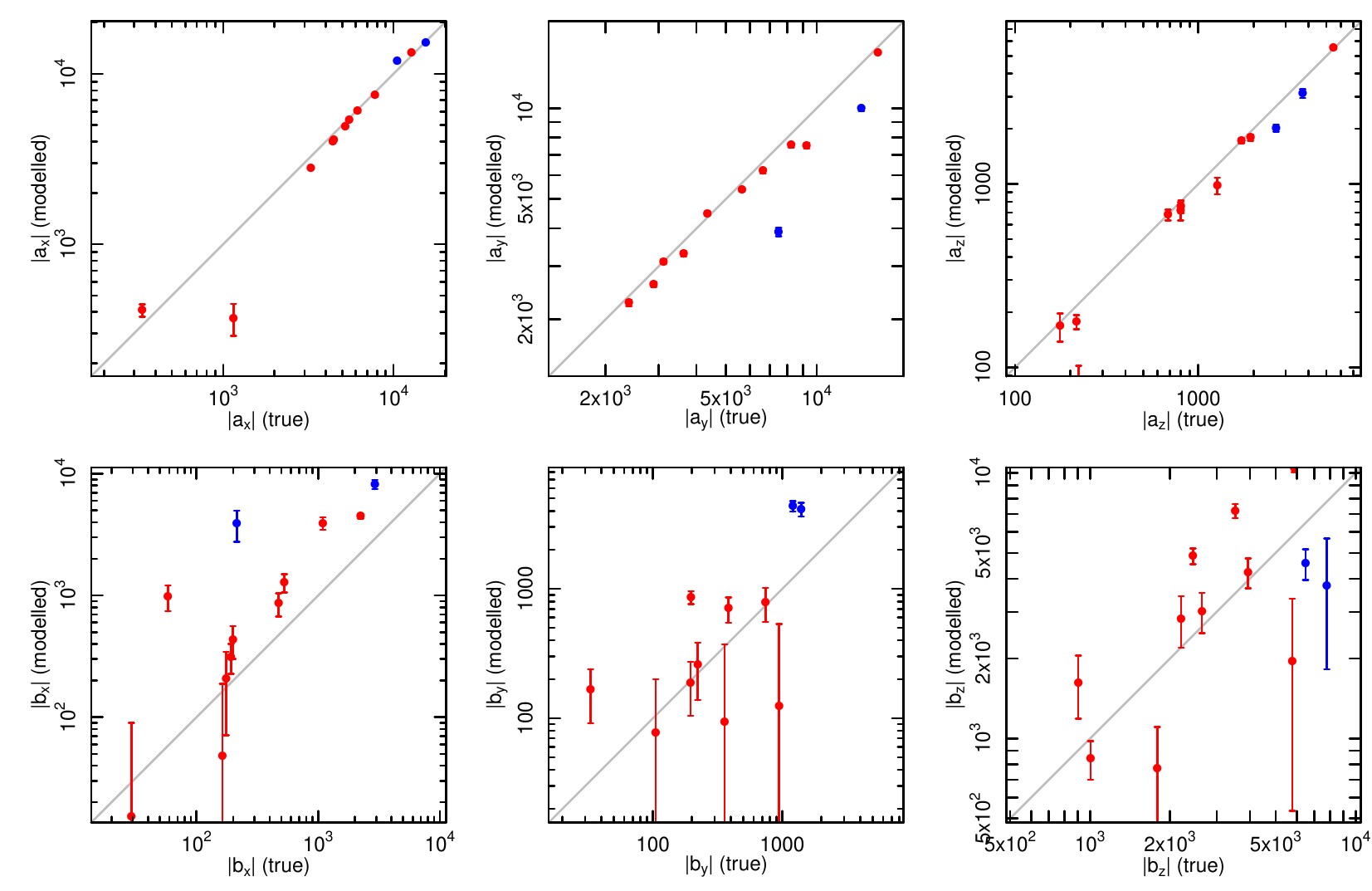}
	\caption{Modelled and true parameters together with error bars of accelerations (see Eq.~(\ref{eq:model_acc})) for twelve regions of the barred snapshot. The errors are marginalised standard deviations of the posterior distribution samples. It is seen that acceleration components are rather well recovered (upper panels), but correction terms for accelerations for several regions have rather large errors. The two outliers (blue dots with their error bars) in $|a_y|$ panel are the ones laying near to the bar ($R1$ and $R4$).} \label{fig:parval}
\end{figure*}

\subsection{Overall recovery of the accelerations}

Calculated and true radial acceleration values $a_R = \sqrt{a_x^2 + a_y^2}$ and their directions for the simple galaxy mock are seen in Figs.~\ref{fig:gal_plane_NE} and \ref{fig:parval_NE}. Lengths of arrows that describe true values were fixed and modelled values were adjusted to those with correct ratios. It is seen that recovery of true accelerations is rather good, calculated $a_R$ values differ from true ones by only $4$ per cent (median). Directions of acceleration vectors differ from true ones by only $1$ degree (median). Individual fitting results are given numerically in Table~\ref{tab:regions}. 

In vertical direction calculated accelerations $a_z$ differ from the true values by $14$ per cent (median). This poorer fit is expected as in case of $z$ direction our used form for $a_z$ (see Eq.~(\ref{eq:model_acc})) is too simple to describe rather quick density changes with $z$. In $z$ direction near to the galactic plane it would be better to use a more complicated form for $a_z (x,y,z)$ with square and even cubic terms and to use thinner regions. However, in this case the time stars spend in these regions decreases also. Increasing the number of stars should compensate this, but unfortunately, we are limited by the available simulation data. We want to point out that the purpose of the present paper is to introduce the method and to make simple tests. Thus, selected form for accelerations should be handled as a first approximation only. In case of more sophisticated modelling the form for acceleration should depend on a specific application and availability and quality of the data. 

Overall, the coefficients ($b_j$) in Eq.~(\ref{eq:model_acc}) contributed about $10$ per cent to the overall acceleration values. Taking into account that within these rather small regions it is not easy to derive accelerations from velocities one may say that derived $b_j$ values are quite uncertain. In principle, increasing the number density of stars in regions should allow to decrease uncertainties of $b_j$, but the number density of stars in the used simulation is limiting that. 

We can notice some correlation (correlation coefficient 0.5) between the relative error of recovered $a_z$ and $a_R$, and between the orientation of the region and underlying rotational direction of stars (correlation coefficient 0.6). This is related with the choice of the box location and geometry and can be taken into account and improved in real applications. 

Real galaxies often include non-axisymmetric components and non-stationarities. To test the method for these galaxies we used the method to model a barred galaxy snapshot. Results of the model calculations are shown in Figs.~\ref{fig:gal_plane} and \ref{fig:parval}, and numerically given in Table~\ref{tab:regions}. Modelling accuracies in this case for $a_R$ and $a_z$ are $5$ and $9$ per cent (median values), respectively. Directions of  acceleration vectors were recovered with less than $2$ degrees (median). 

It is seen from Table~\ref{tab:regions} that three least accurate solutions are for regions \textit{R1}, \textit{R4}, and \textit{R11}. In case of \textit{R1} and \textit{R4}, the model did not reproduced directions of acceleration vectors. This is not surprising, since both regions are near to the bar (see Fig.~\ref{fig:gal_plane}). 

In case of the region \textit{R11} derived acceleration value $a_R$ is quite inaccurate. This is seen also for the simple galaxy snapshot. From simple snapshot results we found a mild correlation ($0.6$) between the position angles of the regions and recovered accuracies -- for regions with best alignment with $x$ or $y$ axis the accuracy of modelled accelerations are the best. This can be interpret in a way that rotating stars move through the corners of these regions very fast and do not sufficiently influence overall acceleration value. Most of the information for modelling is from bins that are near to the centre of the box and are thus derived acceleration values are geometrically biased.

\subsection{Robustness of the modelling}
\begin{figure}
	\includegraphics{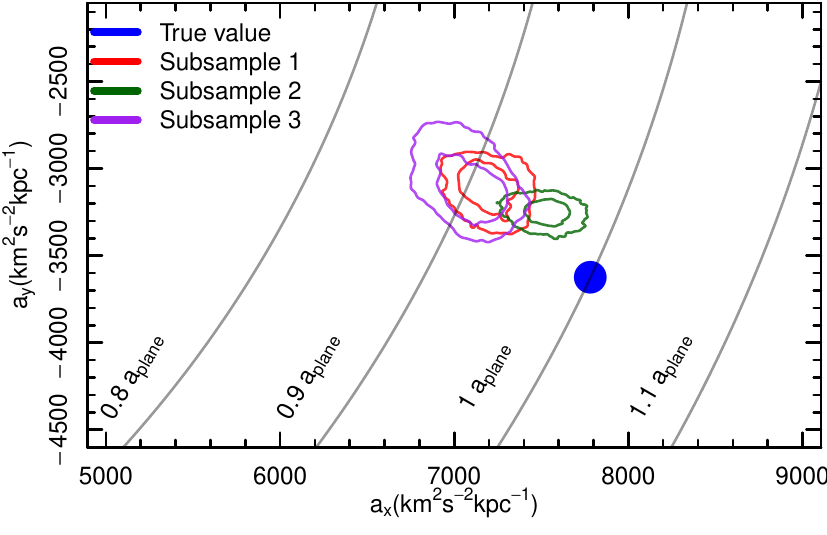}
	\caption{Precision of calculated accelerations in galactic plane for different sampling in case of the region \textit{R10}. Narrow grey lines are curves of constant acceleration in galactic plane with labels indicating the fraction from the true value. The contours show one and two sigma confidence intervals. All the samples are independent and contain equally $38\,297$ stars. }\label{fig:shotnoise}
\end{figure}   
\begin{figure}
	\includegraphics{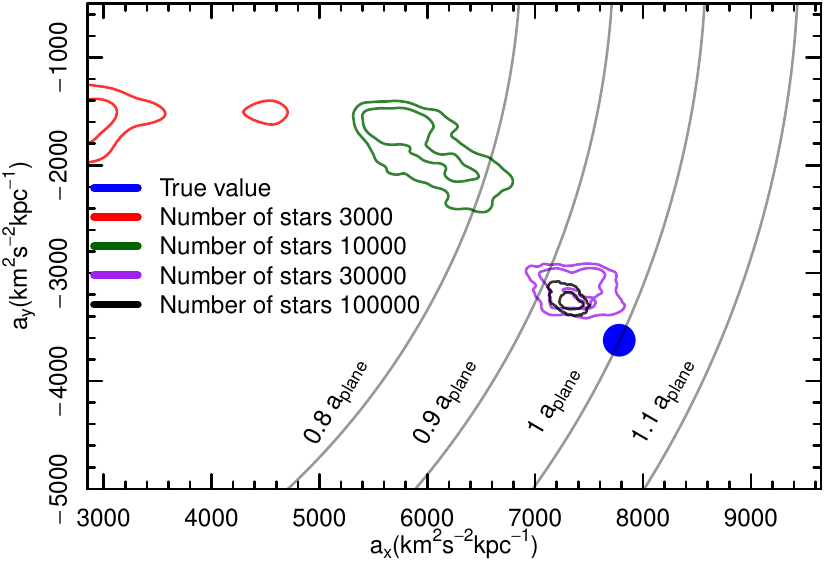}
	\caption{Precision of calculated accelerations in galactic plane as a function of the number of test stars used in modelling in case of the region \textit{R10}. Narrow grey lines are the curves of constant acceleration in galactic plane with labels indicating the fraction from the true value. The contours show one and two sigma confidence intervals. }\label{fig:nstar}
\end{figure}
\begin{figure}
	\includegraphics{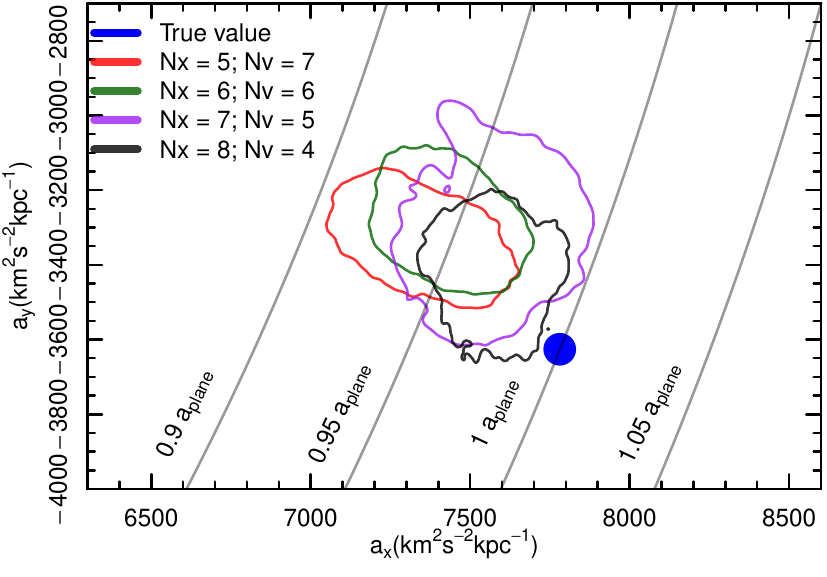}
	\caption{Precision of calculated accelerations in galactic plane for different grids in case of the region \textit{R10}. Narrow grey lines are curves of constant acceleration in galactic plane with labels indicating the fraction from the true value. The contours show two sigma confidence intervals for each grid. Grid sizes are indicated separately for coordinate space (\textit{Nx}) and for velocity space (\textit{Nv}). }\label{fig:ngrid}
\end{figure}
First, we checked whether Multinest fitting converges to the same posterior distribution when running the fitting process with different random seeds. We reran complete fitting process for all twelve regions in case of the barred galaxy snapshot and calculated the differences of the mean values of derived $a_R$ distributions, then divided the result by the averaged standard deviation of the two runs. We found that on average (over all regions) this quantity was $0.5$, the maximum difference was $1.0$. We can conclude that there are some convergence issues, but as we will see, these are smaller than shot noise effects. We conclude that this difference is small enough that it does not change overall our results and conclusions. 

Subsequent tests were done for the region \textit{R10} in the barred galaxy sample. We preferred the barred snapshot over simple one to have more realistic estimate on how well would the model behave in real observations. 

To test for robustness against shot noise, we created three subsamples for \textit{R10}, all of them containing $38\,297$ stars. The results of modelling for each of these subsamples are seen in Fig.~\ref{fig:shotnoise}. The contours show one and two sigma posterior distribution regions. We conclude that the shot noise has larger impact on resulting uncertainties than statistical uncertainties.  

In calculations we used regions containing about $10^5$ test stars (although half of them were actually used) and divided modelled regions into a grid with $6^6$ cells per region. Next we will test how robust these choices are. To test the requirements for the number of test stars, we calculated accelerations by using  different numbers of stars. For this we created from the initial number of $10^5$ stars a random subsets of $3000$, $10\,000$ and $30\,000$ stars. Box sizes remained the same. Fitting results are given in Fig.~\ref{fig:nstar}. We can see that for this region about $30~000$ stars are needed to recover acceleration $a_x$ and $a_y$ components within about 5 per cent. Of course the minimum number of stars depends how smooth is the gravitational potential, are there perturbations in phase density distribution, what is the size and shape of the region etc. This number of stars needed for accurate modelling should be handled as an order of magnitude estimate.

Next we tested sensitivity of the modelling process to grid parameters, more precisely to the number of cells within a given region. Number of test stars remained unchanged (50~000 stars). Four different combinations of numbers for grid coordinates and velocities were used. In all combinations there is about one test star per grid cell. Results of these tests are given in Fig.~\ref{fig:ngrid}. It is seen that there are no systematic trends how the grid is formed as long as there are enough stars to fill the grid.

\section{Summary and discussion} 
\label{sec:summary}
We developed a method to calculate acceleration vectors (gravitational potential gradients) within regions containing tens of thousands of stars with known coordinates and velocities. The method assumes that phases of stars on orbits within a given region are mixed, that is the exact position of a star on its orbit is a random variable with a probability density uniform in time. By knowing all phase space coordinates for stars in a given region at a particular moment, orbits of stars within the region can be calculated for some parametrised form of acceleration. Combining these orbital arcs together we calculate the probability function to find stars at some particular phase space point. Then the likelihood can be calculated using modelled stellar probability distributions with the observed phase space coordinates of the stars. The likelihood (or posterior distribution) has a maximum if the calculated probability distribution and the observed distribution of stars are consistent. This allows to set constraints to the accelerations used in stellar orbit calculations. 

The method assumes that phases of stars are mixed within selected regions where orbits are calculated. In all regions orbits and accelerations are calculated independently to other regions. Thus, there is no need to assume overall stationarity of a galaxy.

We tested the method with two snapshots from a mock galaxy simulation. The first one (denoted as a simple galaxy) corresponds to an axisymmetric galaxy. Twelve regions were selected each containing 100\,000 stars. Method allows to reproduce true radial acceleration values within 4 per cent and directions within 1 degree (median values).

The second snapshot corresponds to the barred galaxy having thus perturbed kinematics. In this case the method allows to reproduce true radial acceleration values within 5 per cent and directions within 2 degrees (median values). A similar representation of potential gradient or acceleration, circular velocity, can be calculated within $3$ per cent and are thus also rather accurate.  

One of the assumptions of the method in its present form was that acceleration in Eq.~(\ref{eq:liikumisvorrandid2}) is a function of coordinates only, i.e. we assume that there are no significant perturbations near and within the studied regions $D$. As the tests for barred galaxy regions \textit{R1} and \textit{R4} showed (Fig.~\ref{fig:gal_plane}), indeed, the model did not reproduce true accelerations well. Our preliminary tests gave that fitting accuracy for these regions can be improved by slightly changing Eq.~(\ref{eq:pi_vorrand}) i.e. allowing mild non-uniformity. 

Proposed method can be used for the Milky Way by using Gaia complete solution data with known spatial coordinates and velocities. By selecting modelled regions $D$ to be sufficiently small the gravitational potential derivatives can be calculated at the whole grid of coordinates covered by Gaia data. A minimum size of a region is constrained by the timing argument (acceleration needs time) and by the statistical argument (large number of stars are needed for an accurate likelihood evaluation). 

By using typical velocities (> 100 $\mathrm{km\,s^{-1}}$) and typical accelerations in the Milky Way galaxy and demanding that velocity changes due to acceleration should exceed, for example five times the velocity measurement error one obtains that in the halo or outer disc region the minimal box size is about $0.5$~kpc. This box contains also sufficient number of stars for statistics. Thus the grid in the MW where it is possible to calculate acceleration components is quite dense. For a typical star it takes around 2--5~Myrs to move through a box with this minimum size. This gives the minimum time-scale how long typical stellar orbits should be integrated. Another constraint is the accuracy of the measured parallaxes: the uncertainty of parallaxes should be smaller than the grid cell size. Although this hinders the applications, the errors of parallaxes are constantly reduced by longer timebase of Gaia and increasing number of observations. 

We presented here only one possible usage of the method. With modifications the model can be used to calculate accelerations close to the Galactic disc plane, to study disturbances near to the bar and in other interesting cases.

\section*{Acknowledgements}
We thank the Referee for useful comments and suggestions that helped to improve the paper substantially. This work was supported by institutional research funding IUT26-2 and IUT40-2  of the Estonian Ministry of Education and Research. We acknowledge the support by the Centre of Excellence ``Dark side of the Universe'' (TK133) and grant MOBTP86 financed by the European Union through the European Regional Development Fund. 
We thank the Gaia Wiki team, who made their simulation data publicly available. 

\bibliographystyle{mnras}
\bibliography{kindyn_paper}

\begin{thebibliography}{}
\makeatletter
\relax
\def\mn@urlcharsother{\let\do\@makeother \do\$\do\&\do\#\do\^\do\_\do\%\do\~}
\def\mn@doi{\begingroup\mn@urlcharsother \@ifnextchar [ {\mn@doi@}
  {\mn@doi@[]}}
\def\mn@doi@[#1]#2{\def\@tempa{#1}\ifx\@tempa\@empty \href
  {http://dx.doi.org/#2} {doi:#2}\else \href {http://dx.doi.org/#2} {#1}\fi
  \endgroup}
\def\mn@eprint#1#2{\mn@eprint@#1:#2::\@nil}
\def\mn@eprint@arXiv#1{\href {http://arxiv.org/abs/#1} {{\tt arXiv:#1}}}
\def\mn@eprint@dblp#1{\href {http://dblp.uni-trier.de/rec/bibtex/#1.xml}
  {dblp:#1}}
\def\mn@eprint@#1:#2:#3:#4\@nil{\def\@tempa {#1}\def\@tempb {#2}\def\@tempc
  {#3}\ifx \@tempc \@empty \let \@tempc \@tempb \let \@tempb \@tempa \fi \ifx
  \@tempb \@empty \def\@tempb {arXiv}\fi \@ifundefined
  {mn@eprint@\@tempb}{\@tempb:\@tempc}{\expandafter \expandafter \csname
  mn@eprint@\@tempb\endcsname \expandafter{\@tempc}}}

\bibitem[\protect\citeauthoryear{{Barone} et~al.,}{{Barone}
  et~al.}{2018}]{Barone:2018}
{Barone} T.~M.,  et~al., 2018, \mn@doi [\apj] {10.3847/1538-4357/aaaf6e}, \href
  {http://adsabs.harvard.edu/abs/2018ApJ...856...64B} {856, 64}

\bibitem[\protect\citeauthoryear{{Binney}}{{Binney}}{2018}]{Binney:2018}
{Binney} J.,  2018, \mn@doi [\mnras] {10.1093/mnras/stx2835}, \href
  {http://adsabs.harvard.edu/abs/2018MNRAS.474.2706B} {474, 2706}

\bibitem[\protect\citeauthoryear{{Binney} \& {McMillan}}{{Binney} \&
  {McMillan}}{2015}]{Binney:2015}
{Binney} J.,  {McMillan} P.~J.,  2015, {TM: Torus Mapper}, Astrophysics Source
  Code Library (\mn@eprint {ascl} {1512.014})

\bibitem[\protect\citeauthoryear{{Binney} \& {McMillan}}{{Binney} \&
  {McMillan}}{2016}]{Binney:2016}
{Binney} J.,  {McMillan} P.~J.,  2016, \mn@doi [\mnras]
  {10.1093/mnras/stv2734}, \href
  {http://adsabs.harvard.edu/abs/2016MNRAS.456.1982B} {456, 1982}

\bibitem[\protect\citeauthoryear{{Binney}, {Davies}  \& {Illingworth}}{{Binney}
  et~al.}{1990}]{Binney:1990}
{Binney} J.~J.,  {Davies} R.~L.,   {Illingworth} G.~D.,  1990, \mn@doi [\apj]
  {10.1086/169169}, \href {http://adsabs.harvard.edu/abs/1990ApJ...361...78B}
  {361, 78}

\bibitem[\protect\citeauthoryear{{Bovy}}{{Bovy}}{2014}]{Bovy:2014}
{Bovy} J.,  2014, \mn@doi [\apj] {10.1088/0004-637X/795/1/95}, \href
  {http://adsabs.harvard.edu/abs/2014ApJ...795...95B} {795, 95}

\bibitem[\protect\citeauthoryear{{Cappellari}}{{Cappellari}}{2008}]{Cappellari:2008}
{Cappellari} M.,  2008, \mn@doi [\mnras] {10.1111/j.1365-2966.2008.13754.x},
  \href {http://adsabs.harvard.edu/abs/2008MNRAS.390...71C} {390, 71}

\bibitem[\protect\citeauthoryear{{Cappellari} et~al.,}{{Cappellari}
  et~al.}{2006}]{Cappellari:2006}
{Cappellari} M.,  et~al., 2006, \mn@doi [\mnras]
  {10.1111/j.1365-2966.2005.09981.x}, \href
  {http://adsabs.harvard.edu/abs/2006MNRAS.366.1126C} {366, 1126}

\bibitem[\protect\citeauthoryear{{Cappellari} et~al.,}{{Cappellari}
  et~al.}{2007}]{Cappellari:2007}
{Cappellari} M.,  et~al., 2007, \mn@doi [\mnras]
  {10.1111/j.1365-2966.2007.11963.x}, \href
  {http://adsabs.harvard.edu/abs/2007MNRAS.379..418C} {379, 418}

\bibitem[\protect\citeauthoryear{{Copin}, {Zhao}  \& {de Zeeuw}}{{Copin}
  et~al.}{2000}]{Copin:2000}
{Copin} Y.,  {Zhao} H.~S.,   {de Zeeuw} P.~T.,  2000, \mn@doi [\mnras]
  {10.1046/j.1365-8711.2000.03827.x}, \href
  {http://adsabs.harvard.edu/abs/2000MNRAS.318..781C} {318, 781}

\bibitem[\protect\citeauthoryear{{Emsellem}, {Monnet}  \& {Bacon}}{{Emsellem}
  et~al.}{1994}]{Emsellem:1994}
{Emsellem} E.,  {Monnet} G.,   {Bacon} R.,  1994, \aap, \href
  {http://adsabs.harvard.edu/abs/1994A%26A...285..723E} {285, 723}

\bibitem[\protect\citeauthoryear{{Feroz} \& {Hobson}}{{Feroz} \&
  {Hobson}}{2008}]{MN1}
{Feroz} F.,  {Hobson} M.~P.,  2008, \mn@doi [\mnras]
  {10.1111/j.1365-2966.2007.12353.x}, \href
  {http://adsabs.harvard.edu/abs/2008MNRAS.384..449F} {384, 449}

\bibitem[\protect\citeauthoryear{{Feroz}, {Hobson}  \& {Bridges}}{{Feroz}
  et~al.}{2009}]{MN2}
{Feroz} F.,  {Hobson} M.~P.,   {Bridges} M.,  2009, \mn@doi [\mnras]
  {10.1111/j.1365-2966.2009.14548.x}, \href
  {http://adsabs.harvard.edu/abs/2009MNRAS.398.1601F} {398, 1601}

\bibitem[\protect\citeauthoryear{{Feroz}, {Hobson}, {Cameron}  \&
  {Pettitt}}{{Feroz} et~al.}{2013}]{MN3}
{Feroz} F.,  {Hobson} M.~P.,  {Cameron} E.,   {Pettitt} A.~N.,  2013, preprint,
  \href {http://adsabs.harvard.edu/abs/2013arXiv1306.2144F} {} (\mn@eprint
  {arXiv} {1306.2144})

\bibitem[\protect\citeauthoryear{{Gaia Collaboration} et~al.,}{{Gaia
  Collaboration} et~al.}{2018}]{Gaia_DR2}
{Gaia Collaboration} et~al., 2018, \mn@doi [\aap]
  {10.1051/0004-6361/201833051}, \href
  {http://adsabs.harvard.edu/abs/2018A%26A...616A...1G} {616, A1}

\bibitem[\protect\citeauthoryear{{Garbari}, {Read}  \& {Lake}}{{Garbari}
  et~al.}{2011}]{Garbari:2011}
{Garbari} S.,  {Read} J.~I.,   {Lake} G.,  2011, \mn@doi [\mnras]
  {10.1111/j.1365-2966.2011.19206.x}, \href
  {http://adsabs.harvard.edu/abs/2011MNRAS.416.2318G} {416, 2318}

\bibitem[\protect\citeauthoryear{{Kalinova}, {van de Ven}, {Lyubenova},
  {Falc{\'o}n-Barroso}, {Colombo}  \& {Rosolowsky}}{{Kalinova}
  et~al.}{2017}]{Kalinova:2017}
{Kalinova} V.,  {van de Ven} G.,  {Lyubenova} M.,  {Falc{\'o}n-Barroso} J.,
  {Colombo} D.,   {Rosolowsky} E.,  2017, \mn@doi [\mnras]
  {10.1093/mnras/stw2448}, \href
  {http://adsabs.harvard.edu/abs/2017MNRAS.464.1903K} {464, 1903}

\bibitem[\protect\citeauthoryear{{Kipper}, {Tempel}  \& {Tenjes}}{{Kipper}
  et~al.}{2018}]{Kipper:2018}
{Kipper} R.,  {Tempel} E.,   {Tenjes} P.,  2018, \mn@doi [\mnras]
  {10.1093/mnras/stx2441}, \href
  {http://adsabs.harvard.edu/abs/2018MNRAS.473.2188K} {473, 2188}

\bibitem[\protect\citeauthoryear{{Kowalczyk}, {{\L}okas}  \&
  {Valluri}}{{Kowalczyk} et~al.}{2017}]{Kowalczyk:2017}
{Kowalczyk} K.,  {{\L}okas} E.~L.,   {Valluri} M.,  2017, \mn@doi [\mnras]
  {10.1093/mnras/stx1520}, \href
  {http://adsabs.harvard.edu/abs/2017MNRAS.470.3959K} {470, 3959}

\bibitem[\protect\citeauthoryear{{Li} et~al.,}{{Li} et~al.}{2018}]{Li:2018}
{Li} H.,  et~al., 2018, \mn@doi [\mnras] {10.1093/mnras/sty334}, \href
  {http://adsabs.harvard.edu/abs/2018MNRAS.476.1765L} {476, 1765}

\bibitem[\protect\citeauthoryear{{Long} \& {Mao}}{{Long} \&
  {Mao}}{2010}]{Long:2010}
{Long} R.~J.,  {Mao} S.,  2010, \mn@doi [\mnras]
  {10.1111/j.1365-2966.2010.16438.x}, \href
  {http://adsabs.harvard.edu/abs/2010MNRAS.405..301L} {405, 301}

\bibitem[\protect\citeauthoryear{{McGaugh}, {Lelli}  \& {Schombert}}{{McGaugh}
  et~al.}{2016}]{Mcgaugh:2016}
{McGaugh} S.~S.,  {Lelli} F.,   {Schombert} J.~M.,  2016, \mn@doi [Physical
  Review Letters] {10.1103/PhysRevLett.117.201101}, \href
  {http://adsabs.harvard.edu/abs/2016PhRvL.117t1101M} {117, 201101}

\bibitem[\protect\citeauthoryear{{McGill} \& {Binney}}{{McGill} \&
  {Binney}}{1990}]{McGill:1990}
{McGill} C.,  {Binney} J.,  1990, \mnras, \href
  {http://adsabs.harvard.edu/abs/1990MNRAS.244..634M} {244, 634}

\bibitem[\protect\citeauthoryear{{Schwarzschild}}{{Schwarzschild}}{1979}]{Schwarzschild:1979}
{Schwarzschild} M.,  1979, \mn@doi [\apj] {10.1086/157282}, \href
  {http://adsabs.harvard.edu/abs/1979ApJ...232..236S} {232, 236}

\bibitem[\protect\citeauthoryear{{Syer} \& {Tremaine}}{{Syer} \&
  {Tremaine}}{1996}]{Syer:1996}
{Syer} D.,  {Tremaine} S.,  1996, \mn@doi [\mnras] {10.1093/mnras/282.1.223},
  \href {http://adsabs.harvard.edu/abs/1996MNRAS.282..223S} {282, 223}

\bibitem[\protect\citeauthoryear{{Thomas}, {Saglia}, {Bender}, {Thomas},
  {Gebhardt}, {Magorrian}  \& {Richstone}}{{Thomas} et~al.}{2004}]{Thomas:2004}
{Thomas} J.,  {Saglia} R.~P.,  {Bender} R.,  {Thomas} D.,  {Gebhardt} K.,
  {Magorrian} J.,   {Richstone} D.,  2004, \mn@doi [\mnras]
  {10.1111/j.1365-2966.2004.08072.x}, \href
  {http://adsabs.harvard.edu/abs/2004MNRAS.353..391T} {353, 391}

\bibitem[\protect\citeauthoryear{{Valluri}, {Merritt}  \& {Emsellem}}{{Valluri}
  et~al.}{2004}]{Valluri:2004}
{Valluri} M.,  {Merritt} D.,   {Emsellem} E.,  2004, \mn@doi [\apj]
  {10.1086/380896}, \href {http://adsabs.harvard.edu/abs/2004ApJ...602...66V}
  {602, 66}

\bibitem[\protect\citeauthoryear{{Zhu} et~al.,}{{Zhu} et~al.}{2014}]{Zhu:2014}
{Zhu} L.,  et~al., 2014, \mn@doi [\apj] {10.1088/0004-637X/792/1/59}, \href
  {http://adsabs.harvard.edu/abs/2014ApJ...792...59Z} {792, 59}

\bibitem[\protect\citeauthoryear{{Zhu} et~al.,}{{Zhu} et~al.}{2018}]{Zhu:2018}
{Zhu} L.,  et~al., 2018, \mn@doi [Nature Astronomy]
  {10.1038/s41550-017-0348-1}, \href
  {http://adsabs.harvard.edu/abs/2018NatAs...2..233Z} {2, 233}

\bibitem[\protect\citeauthoryear{{de Lorenzi}, {Debattista}, {Gerhard}  \&
  {Sambhus}}{{de Lorenzi} et~al.}{2007}]{deLorenzi:2007}
{de Lorenzi} F.,  {Debattista} V.~P.,  {Gerhard} O.,   {Sambhus} N.,  2007,
  \mn@doi [\mnras] {10.1111/j.1365-2966.2007.11434.x}, \href
  {http://adsabs.harvard.edu/abs/2007MNRAS.376...71D} {376, 71}

\bibitem[\protect\citeauthoryear{{van de Ven}, {de Zeeuw}  \& {van den
  Bosch}}{{van de Ven} et~al.}{2008}]{vdven:2008}
{van de Ven} G.,  {de Zeeuw} P.~T.,   {van den Bosch} R.~C.~E.,  2008, \mn@doi
  [\mnras] {10.1111/j.1365-2966.2008.12873.x}, \href
  {http://adsabs.harvard.edu/abs/2008MNRAS.385..614V} {385, 614}

\bibitem[\protect\citeauthoryear{{van der Marel}, {Binney}  \& {Davies}}{{van
  der Marel} et~al.}{1990}]{vdMarel:1990}
{van der Marel} R.~P.,  {Binney} J.,   {Davies} R.~L.,  1990, \mnras, \href
  {http://adsabs.harvard.edu/abs/1990MNRAS.245..582V} {245, 582}

\makeatother
\end{thebibliography}
\label{lastpage}
\end{document}